\documentclass[final,3p,times,twocolumn]{elsarticle}

\usepackage{amssymb}
\usepackage{amsthm}
\usepackage{amsmath,graphicx}
\usepackage{subcaption}
\usepackage{multirow}
\usepackage{url}

\journal{Nuclear Physics B}

\begin{document}
\begin{frontmatter}

\title{Distilling Multi-Level X-vector Knowledge for Small-footprint Speaker Verification}

\author[inst1]{Xuechen Liu}
\author[inst3]{Md Sahidullah}
\author[inst2]{Tomi Kinnunen}

\address[inst1]{National Institute of Informatics, 2-1-2 Hitotsubashi, 101-8430 Tokyo, Japan}
\address[inst2]{University of Eastern Finland, L\"{a}nsikatu 15, FI-80110 Joensuu, Finland}
\address[inst3]{Institute for Advancing Intelligence, TCG CREST, 700091 Kolkata, India}

\begin{abstract}
Even though deep speaker models have demonstrated impressive accuracy in speaker verification tasks, this often comes at the expense of increased model size and computation time, presenting challenges for deployment in resource-constrained environments. Our research focuses on addressing this limitation through the development of small-footprint deep speaker embedding extraction using knowledge distillation. While previous work in this domain has concentrated on speaker embedding extraction at the utterance level, our approach involves amalgamating embeddings from different levels of the x-vector model (teacher network) to train a compact student network. The results highlight the significance of frame-level information, with the student models exhibiting a remarkable size reduction of 85\%-91\% compared to their teacher counterparts, depending on the size of the teacher embeddings. Notably, by concatenating teacher embeddings, we achieve student networks that maintain comparable performance to the teacher while enjoying a substantial 75\% reduction in model size. These findings and insights extend to other x-vector variants, underscoring the broad applicability of our approach.
\end{abstract}




\end{frontmatter}


\section{Introduction}
\label{sec:intro}
\emph{Automatic speaker verification} (ASV) \cite{dnn_asv2021} aims at identifying individuals using their voices. Advancing upon traditional, shallow statistical comparators \cite{simiarity_asv2010}, \emph{Gaussian mixture models} (GMM) \cite{gmm,gmm_earlyasv_2008} and \emph{i-vectors} \cite{ivector,ivector_asv2019}, \emph{deep neural networks} (DNNs) have emerged as the modern ASV solution \cite{dnn_asv2021}. Currently, DNNs are the predominant choice for representing speakers using \emph{speaker embeddings} --- fixed-dimensional vectors containing speaker-specific traits.

Meanwhile, proliferating applications of speech processing and biometric algorithms onto embedded devices \cite{assistant1,assistant2,assistant3} with constrained local computational resources have emerged. Representative examples are smart home speakers such as Amazon Alexa\footnote{https://developer.amazon.com/en-US/alexa} and Google Home\footnote{https://home.google.com/}, and assistance devices such as portable voice translator\footnote{https://global.iflytek.com/jarvisen/}. DNNs have various constraints to their practical deployment on them, including run-time (in)efficiency, power consumption, and memory usage due to a large number of parameters \cite{kd2006}. This is a major shortcoming when the embedded device has limited memory space or needs to operate with weak or no online access, ruling out cloud-based solutions. Therefore, lightweight on-device models are preferable --- though this usually comes with a trade-off in recognition accuracy. Reducing the performance gap between the small and large DNN models is an important, yet challenging task in speech processing \cite{ondevice3,ondevice1,tslearning_slt2018}. 

Therefore, in this paper, we aim at reducing this gap to achieve acceptable trade-off between the model size and performance. In particular, we propose a 
simple and practical approach for creating a small `student' DNN from a larger `teacher' counterpart through \emph{knowledge distillation} (KD). We review the relevant related work on small footprint ASV and KD in Sections 
\ref{sec:background} and \ref{sec:methodology}, respectively. While our focus is on ASV using \emph{de facto} off-the-shelf speaker encoders,
the proposed distillation framework can readily be extended to other networks and tasks (as will be discussed in Section \ref{ssec:configs}). 

While KD itself is not new to speech processing tasks \cite{kd_asr2013,kd_asr2018,kd2021}, the practical implementation details are crucial for achieving desired model size--accuracy trade-off for a given model and task. We put forward a novel perspective on \emph{multi-level} speaker representation suited for small footprint ASV. Given that the multiple layers in a DNN represent different types of features---ranging from low-level short-term spectral descriptors to abstract utterance-level latent speaker traits---a key question concerns the usefulness of the information represented by the different layers of the teacher network. \textbf{While the common approach is to extract a single, utterance-level speaker embedding, we hypothesize that embeddings extracted from \emph{all} the layers could help in reducing the performance gap of the teacher and the student networks.} 
After formulating our approach in Section \ref{sec:embeddings}, we seek to confirm the above hypothesis experimentally in Section \ref{sec:experiments}. 
To this end, we investigate the effect of embedding-level composite (using simple concatenation) and extend the proposed methods to more advanced x-vector variants including the ones described in \cite{Snyder_etdnn_2019,ecapa,dtdnn}. \textbf{To the best of our knowledge, this is the first work that reveals the efficacy of an `extended' set of multi-level speaker embeddings for developing practical, small footprint ASV systems.}

\begin{figure*}[t]
    \centering
    \includegraphics[width=0.8\linewidth]{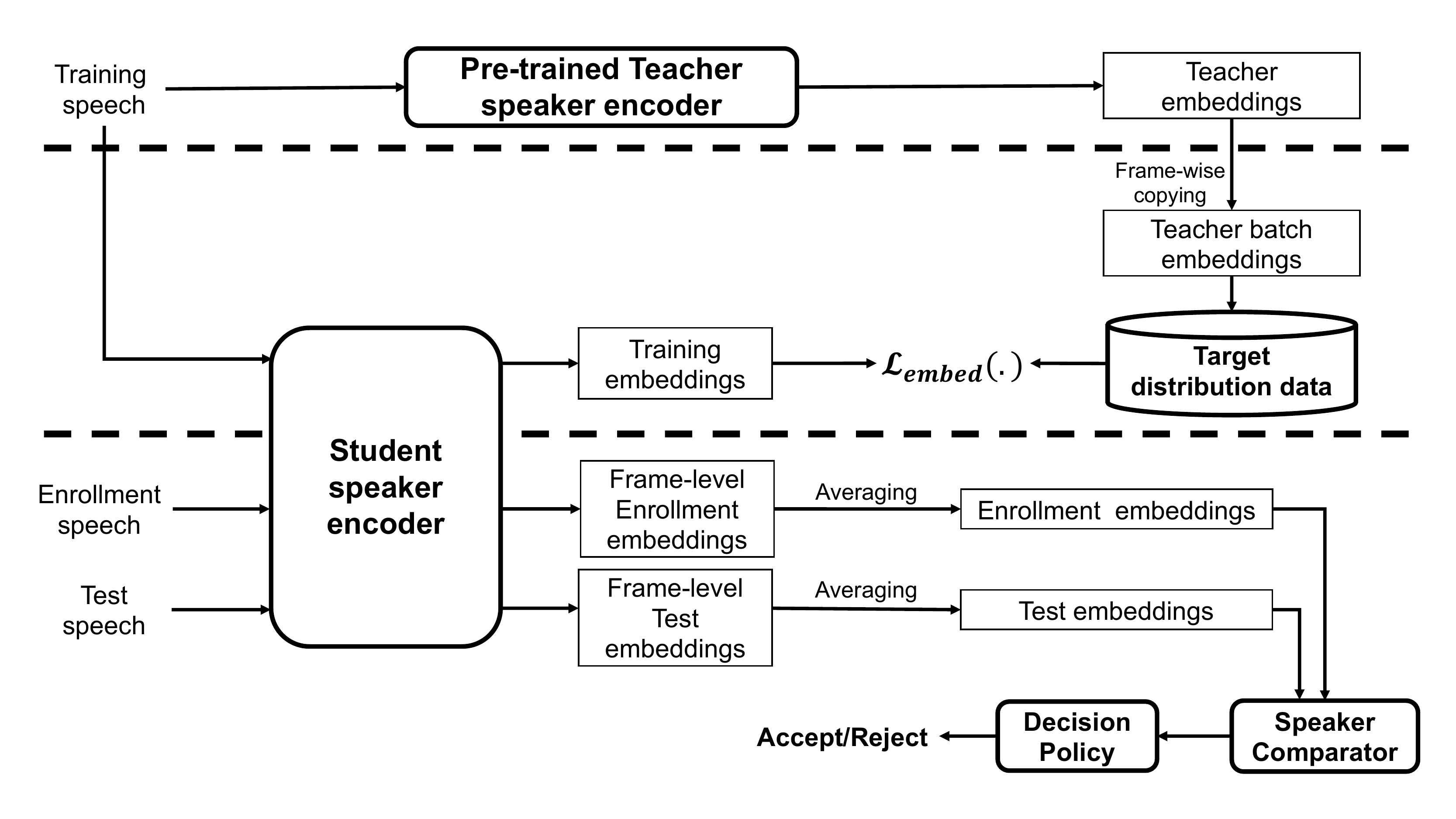}
    \caption{The TS learning framework used for this study. The pre-trained teacher speaker encoder reads training speech waveforms and outputs speaker embeddings of them at the first stage (up, divided by the thick dash lines). The training utterances are then fed into the student speaker encoder and the learning proceeds by comparing the student prediction and the frame-wise copied speaker embeddings at the student training stage (middle). At the run-time stage (bottom), the frame-level enrollment and test embeddings are extracted and averaged to single embeddings, for decision making and evaluation.}
\label{fig:tslearning}
\end{figure*}

\section{Small Footprint Speaker Verification}
\label{sec:background}
Modern neural networks for ASV are typically trained on thousands of speakers and comprise millions of learnable parameters. Despite impressive performance on common ASV evaluation corpora, the size and complexity of these models pose deployment challenges on embedded devices with limited memory and computational power. This challenge is highlighted not only for training but also, more importantly, during inference, which makes modern ASV systems be sometimes accessible only through cloud-based services with strong hardware backbones. In environments with weak or non-existent internet connection, a preference may arise for more lightweight variants of these systems, thereby alleviating the constraints on processing power and memory resources. Meanwhile, there has been a consistently-observed trade-off between model complexity and performance \cite{energy_efficient_net2018,tslearning_slt2018}. Therefore, it becomes important to ensure that down-scaled models maintain performance levels comparable to their larger counterparts. 

Solutions to this problem can be divided into two broad categories. The first one, \emph{model down-scaling}, includes training more compact networks \cite{mobilenet,efficientnet}, or quantizing the model \cite{quantized_net}. For instance, \cite{binarized_net_asv2021} proposed parameter binarization for ASV models and \cite{audio_albert} reduced the size of self-supervised speech models via parameter sharing. While achieving promising performance in particular tasks, these approaches may demand extensive engineering effort and subtle learning schemes such as model quantization and parameter tuning. The second category, \emph{knowledge distillation} (KD), also known as \emph{teacher-student} (TS) learning \cite{kd2006,kd2021,kd2022}, transfers the knowledge from a \emph{teacher} network (usually a pre-trained DNN) to a new, \emph{student} network. Compared to the first category, KD possesses task-specific design of student network and knowledge \cite{tslearning_slt2018,kd2020} and tends to require less hyperparameter tuning --- especially when a pre-trained teacher model is readily available (e.g. \cite{ecapa}\footnote{\url{https://huggingface.co/speechbrain/spkrec-ecapa-voxceleb}}, \cite{dtdnn}\footnote{\url{https://github.com/yuyq96/D-TDNN}}). Nevertheless, producing effective knowledge for student learning remains challenging \cite{kd2021}, especially when the student network is expected to be significantly smaller than the teacher. 

Having been convinced on the overall positive prospects and  simplicity of the generic KD framework (stemming partly from earlier hands-on experience of the first author \cite{kd_asv2018}), our work seeks to develop a novel, small footprint ASV model using KD. When DNNs were initially used for speaker embedding extraction, an open question was how the embedding should be extracted \cite{xvector_2016} --- in particular, \emph{which DNN layer(s) would be most useful for speaker characterization}. Models such as an x-vector with \emph{time-delay neural network} (TDNN) use utterance-level (referred to as `segment-level' in \cite{xvector2017}) speaker embeddings, which contains abstract high-level latent speaker information. Meanwhile, the frame-level information at earlier layers can still be useful and its successful application as bottleneck features has been studied \cite{dvector,bnf}, which motivates part of the methodology of this work.

There has been two major approaches towards producing the small footprint neural networks from their large-scale counterparts, which are detailed below.

\subsection{Model Quantization}
A pragmatic approach to reduce the size of a large DNN to deployable sizes involves quantizing the model itself. This option is not only practical but also reasonable, given the redundancies inherent in DNNs during training, including ineffective connections and void parameters. Recent efforts, including those focused on state-of-the-art DNN speaker encoders \cite{ecapa}, have delved into visualizing receptive feature maps from various layers to understand their functions. 

One strategy to quantize the network is to prune the parameters. Noteworthy contributions in this domain, as exemplified by \cite{deep_compression,deep_compression2015}, entail the removal of weights and associated connections through either elementary pruning followed by subsequent quantization and Huffman encoding, or the application of weight clustering into discrete groups represented by centroids. The latter methodology, through the transformation of centroid values, facilitates resource-efficient weight sharing while concurrently mitigating network sparsity. In the realm of \emph{convolutional neural networks} (CNN), the implementation of pruning across specific dimensions within the convolutional layers has been explored \cite{cnn_compression2017}. Concurrently, studies have been conducted to investigate error minimization in this context \cite{cnn_compression2016}. A recent, promising approach involves binary (1-bit) quantization of weights \cite{binarized_nnet2021}, applied to speaker embedding extraction through conventional \emph{time-delayed neural networks} (TDNNs) \cite{binarized_nnet2020}. Alternatively, eliminating redundant parameters can be done via \emph{low-rank parameter factorization} \cite{cnn_decomposition2015} achieved by estimating the importance of weight parameters through matrix decomposition techniques.

Additionally, enhancing the processing efficiency of convolutional filter kernels to decrease computational complexity is another viable option. This can be achieved through factorizing the convolution kernel to two separate, cascaded kernels, with significant size reduction \cite{mobilenet_v2,mobilenet_asv2020}. Another strategy involves splitting the filters across different channels, followed by shuffling and re-grouping, a methodology illustrated in works such as \cite{shufflenet,shufflenet_v2}.

\subsection{Knowledge Distillation}
When implementing knowledge distillation, a certain level of information is extracted from a large ensemble of models as pseudo data distribution, which is then utilized to train a DNN model. The distribution can be generated from unlabeled data using various methods, including random sampling, mixture modeling, and designated algorithms \cite{kd2006}.

Originally applied to speech processing tasks \cite{kd_asr2013,kd_asr2018}, KD served primarily as a means to effectively transfer the knowledge of a well-trained network to a new one---but without necessarily prioritizing model size reduction. For instance, in \cite{kd_asr2013}, KD was employed as a domain adaptation technique, where the training data for the teacher network constituted the source domain, and the training data for the student network constituted the target domain. This approach is further explored in \cite{kd_asr2018}, where a semi-supervised objective based on lattice-free maximum mutual information was proposed. Both studies concentrate on speech recognition, with the training target being the state estimates of the \emph{Hidden Markov model} (HMM), typically designed as the training target for the acoustic model.

\section{Knowledge Distillation for Speaker Verification}
\label{sec:methodology}
In this section, we outline our methodology of developing and training a streamlined DNN speaker encoder. We use a pre-trained DNN model as teacher and implement Knowledge Distillation (KD) via a simple student network. We begin by providing details about our pre-trained teacher model and then proceed to discuss the configuration of our learning framework.

\subsection{X-Vector}
\label{ssec:xvector}
\begin{figure}[t]
    \centering
    \includegraphics[width=0.9\linewidth]{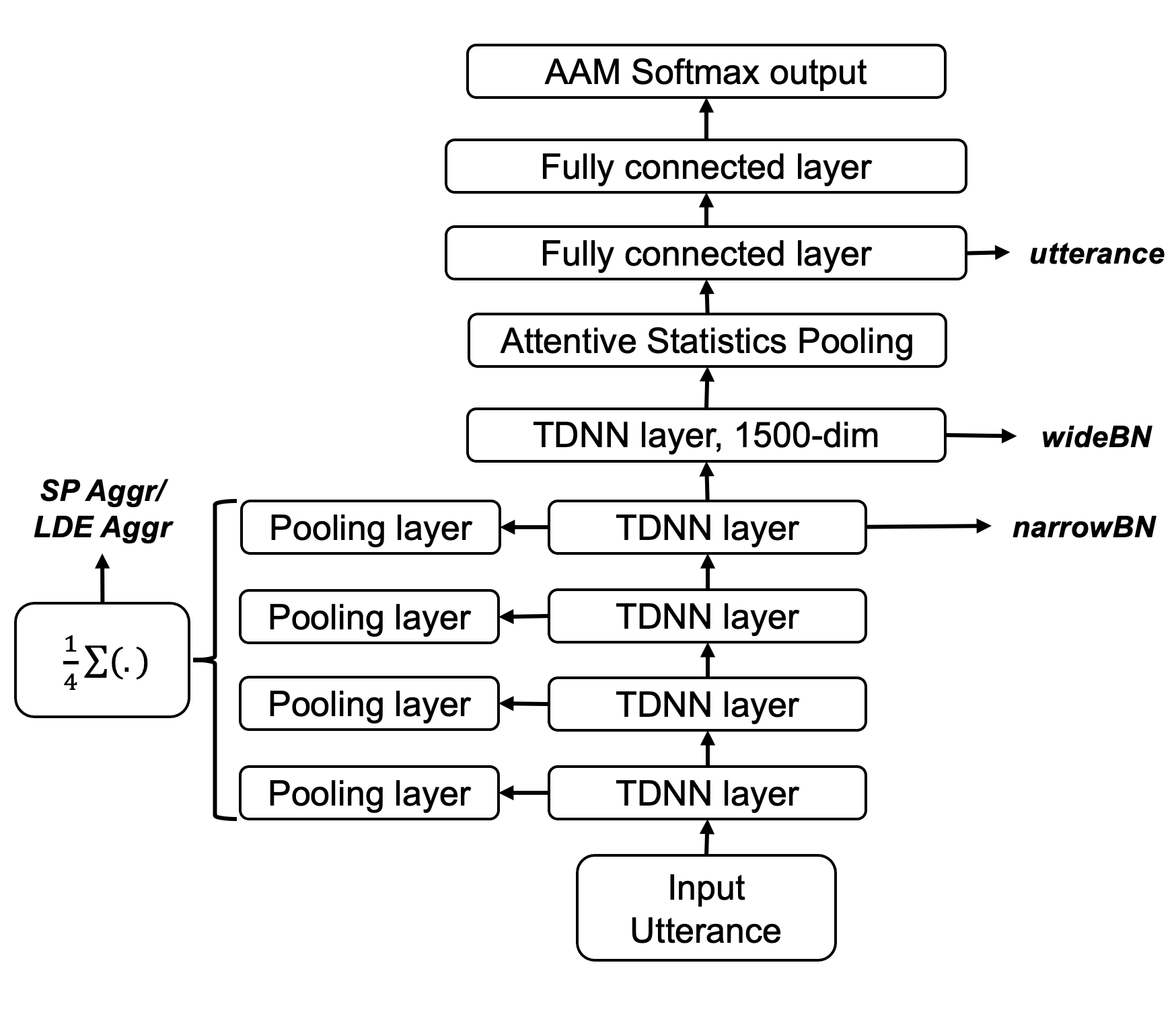}
    \caption{The x-vector model and the various types of embedding that are extracted from its layers. The TDNN and fully connected layers, if not marked, are all 512-dimensional. `AAM softmax' denotes \emph{additive angular margin softmax} \cite{aam_softmax}. The pooling layer on the left-hand side can be either statistics pooling or learnable dictionary encoding (LDE).}
\label{fig:xvector}
\end{figure}

Given its efficacy and transparency among existing \emph{deep neural network} (DNN) models \cite{dnn_asv2021}, we consider the widely recognized vanilla x-vector speaker encoder \cite{xvector2017,xvector2018}. This model is founded on the \emph{time-delay} neural network (TDNN) architecture \cite{tdnn2015}, as depicted in Fig. \ref{fig:xvector}. Selected for its efficiency in terms of model parameters, the vanilla x-vector model comprises five frame-level TDNN layers, a subsequent pooling layer, two utterance-level fully-connected layers, and an output layer. This choice allows us to explore the nuances of minimal model parameter compression using a common student network. Following the common practice, compared to the original vanilla x-vector, we substitute statistical pooling with \emph{attentive statistics pooling} \cite{astats_pooling}, and employ \emph{additive angular margin softmax} (AAM softmax, as depicted in Fig. \ref{fig:xvector}) \cite{aam_softmax} as the loss function, deviating from the conventional softmax cross-entropy. While we are aware of various data augmentation techniques \cite{rir,musan} and more advanced models \cite{ecapa,dtdnn} (addressed in section \ref{sec:results}), we choose not to factor in the combination of the influence of data, teacher model complexity, and the effectiveness of the proposed simple learning framework in the main body of this study.

\subsection{The Learning Framework of Knowledge Distillation}
\label{ssec:kd}
Originating in \cite{kd2006} and further explored in \cite{kd2015}, knowledge distillation has recently gathered significant attention. This technique harnesses the robust representation capabilities of a well-trained, large-scale neural network to impart knowledge to a smaller counterpart. In speech and speaker recognition, the teacher network is typically trained using standard objectives like cross-entropy on a massive training corpus. Subsequently, the student network assimilates knowledge by extracting the ``target distribution data" from the teacher \cite{tslearning_slt2018,ts_asr_slt2018,kd2020}, as illustrated in the upper part of Fig. \ref{fig:tslearning}. The primary objective is to align this distribution closely with the task-specific target distribution. In the case of ASV, the latter involves either speaker posterior probabilities or the direct use of speaker embedding vectors. We categorize these as \emph{posterior-} and \emph{embedding-level} information, respectively.

We integrate embedding-level KD, following the design outlined in \cite{ts_asr_slt2018} and \cite{tslearning2019}, with slight adjustments tailored to our systems. The framework is illustrated in the lower part of Fig.~\ref{fig:tslearning}. Informed by preliminary experiments, we opt for cosine similarity \cite{cosine_similarity} as the designated loss function, defined as follows:

\begin{align}
    \ell_{\mathrm{emb}}(\boldsymbol{x}_{t}, \boldsymbol{x}_{s}) = -\sum_{i=1}^{N} \frac{\boldsymbol{x}_{t}^{i} \cdot \boldsymbol{x}_{s}^{i}}{\lVert\boldsymbol{x}_{t}^{i}\rVert \lVert\boldsymbol{x}_{t}^{i}\rVert}
\label{eq:soft}
\end{align}
where $\boldsymbol{x}_{t}^{i}$ and $\boldsymbol{x}_{s}^{i}$ are teacher and student embeddings from the $i$-th sample, respectively. $N$ denotes the number of samples. The loss is computed in a per-frame manner and the returned values are summed and averaged: $\ell_\mathrm{frame\_emb} = \frac{1}{T}\sum_{t=1}^{T} \ell_{\mathrm{emb}}$, where $t$ denotes the frame index and $T$ denotes the number of total frames.

\section{Embeddings for Knowledge Distillation}
\label{sec:embeddings}
While the use of multiple levels of speaker information \cite{aggregated_stats_2020} and KD \cite{kd2020,kd2022} have been addressed in prior work, their combination for small footprint ASV demands a further investigation. 

In this section, we present our methodology by examining different types of speaker embeddings for training student networks. The speaker embeddings are derived exclusively from the teacher network, as illustrated in Fig. \ref{fig:xvector}. Notably, these design choices can be seamlessly transposed onto various other DNN architectures for ASV \cite{dnn_asv2021}.

\subsection{Utterance-level Embeddings}
The prevalent type of speaker embedding in deep speaker verification systems is utterance-level embedding \cite{xvector2017,xvector2018}. For the x-vector model, we extract these embeddings, one per utterance, from the output of either the first or the second fully-connected layer following the statistics pooling layer. As indicated by \cite{xvector2018} and confirmed by our preliminary experiments, there is no substantial performance difference between extracting from the first or the second layer. We opt for the former and term these embeddings as \emph{utterance} in the experimental section.

To synchronize various levels of information, we adhere to the approach described in \cite{tslearning_slt2018} through the following steps: 1) During the training of the student network, we replicate the speaker embedding $T$ times to generate frame-level target distribution ($T$ is set to 2 seconds for all experiments); 2) When extracting embeddings from the student network, we compute the average of the frame-level information across the input utterance.

\subsection{Frame-level Embeddings}
In the x-vector, the layers preceding the statistics pooling step encapsulate various \emph{frame-level} information, while the subsequent layers aggregate this information into an utterance-level representation \cite{xvector2017}. Despite the prevalence and popularity of x-vector variants, where utterance-level embeddings are the norm, the potential of \emph{frame-level} information at early layers has been relatively overlooked. Motivated by earlier investigations into bottleneck (BN) features \cite{dvector,bnf}, we explore the efficacy of \emph{frame-level} embeddings as the distribution for student network learning. Specifically, we investigate bottleneck embeddings extracted from the last two TDNN layers of the teacher network, with the embeddings obtained by averaging over frame samples. The two types of embeddings are as follows:
\begin{itemize}
    \item The 4-th TDNN layer. We refer to this type of embedding as \emph{narrowBN}. The dimension of the vector is 512.
    \item The 5-th TDNN layer, which is the final layer before the attentive statistics pooling. We refer to this type of embedding as \emph{wideBN}. The dimension of the vector is 1500.
\end{itemize}
These types of embeddings share similarities with those from models such as \emph{d-vector} \cite{dvector}. The difference lies in the fact that, in this context, since the teacher model is an x-vector, during teacher training, utterance-level knowledge is back-propagated to the frame-level layers.

Additionally, for these frame-level embeddings, we explore the impact of mean normalization \cite{cmvn}. Specifically, we implement windowed mean normalization with a window size of $s=300$ frames and observe its normalization effect on the embeddings.

\subsection{Aggregated Embeddings}
Aiming to further exploit the potential of frame-level representations, drawing inspiration from \cite{aggregated_stats_2020}, we adopt a strategy that involves aggregating the output from multiple frame-level layers to utterance-level. The resulting embedding is obtained by taking the equal-weighted average of these aggregated outputs. Denoting $\boldsymbol{x}_{t}$ as the output teacher embedding, this approach is expressed as follows:
\begin{align}
    \boldsymbol{x}_{t} = \frac{1}{K}\sum_{i=1}^{K}\mathbf{\Phi}(\boldsymbol{x}_{i}),
\end{align}
where $\mathbf{\Phi}(.)$ denotes the pooling operator and $\boldsymbol{x}_{i}$ is the output from the $i$-th TDNN layer. $K$ is the number of vectors to be aggregated, which was set to be 4 here since we aggregate the first 4 TDNN layers (out of 5 in vanilla x-vector).

We benchmark the efficacy of two variants for $\mathbf{\Phi}(.)$. The first operator is \emph{statistics pooling} (SP) \cite{xvector2017}, wherein the mean and standard deviation of $\boldsymbol{x_{i}}$ are computed and concatenated. The second operator is \emph{learnable dictionary encoding} (LDE) \cite{lde2018}, involving the aggregation of frame-level information from the TDNN layers through a convolutional and a linear encoding layer. The initialization of related weight values in LDE is accomplished using a normal distribution \cite{normal_distribution_2010}. The corresponding embeddings derived from these operators are denoted as \emph{SP Aggr} and \emph{LDE Aggr}, respectively.

\section{Experimental setup}
\label{sec:experiments}

\subsection{Data}
\label{ssec:data}
All experiments are conducted on the VoxCeleb dataset \cite{voxceleb1,voxceleb2}, consisting of two subsets: VoxCeleb1 with 1251 speakers and VoxCeleb2 with 6114 speakers. The training partition of VoxCeleb1, comprising 1211 speakers, and the \emph{dev} set of VoxCeleb2, with 5994 speakers, are utilized for training our speaker encoders and backend classifiers. The performance of all systems is evaluated on the \emph{VoxCeleb1 test set} \cite{voxceleb1}, a dataset encompassing 40 speakers, 4874 utterances, and 37720 trials. The acoustic features employed are 40-dimensional Mel filterbank coefficients with a sampling rate of 16~kHz.

\subsection{Experimental Configuration}
\label{ssec:configs}
\textbf{Baseline teacher model}. For the baseline vanilla x-vector, we follow the configuration outlined in Section~\ref{ssec:xvector}. This x-vector model is trained using the training set of VoxCeleb1 and the \emph{dev} set of VoxCeleb2, encompassing a total of 7205 speakers.

\textbf{Student models}. All student models employ an 8-layer fully-connected Deep Neural Network (DNN) without pooling layers or residual connectors. Each fully-connected layer has a dimensionality of 256, except for the final layer, the dimension of which is determined by the teacher embeddings. Consequently, the size of the models ranges from 0.54M to 0.81M, as detailed in Table~\ref{tab:student_results}.
Compared to the teacher, the single student models reduce the effective size of the model from 85\% to 91\%. The teacher embeddings $\boldsymbol{x}_{t}$ in Eq. (\ref{eq:soft}) are derived from the x-vector using the aforementioned methods. For each system, the speaker embeddings undergo centering and mean normalization before being employed to train a \emph{probabilistic LDA} (PLDA) classifier \cite{kaldi_aplda}. This classifier is then utilized to generate log-likelihood scores for evaluation. 
The \emph{dev} set of VoxCeleb2 is used to train the student models. Both the teacher and student speaker encoders are implemented using PyTorch \cite{pytorch}, and the other modules of the learning framework are implemented using Kaldi \cite{kaldi}.

\textbf{Composite student model}. We first conduct an embedding-level composition by concatenating the diverse types of speaker embeddings obtained from the teacher model. Subsequent steps for training and evaluation mirror those employed for student models trained on individual types of embeddings. The reported results encompass scenarios where the frame-level embeddings from the teacher model are both mean-normalized and not. We note this system as \emph{Composite} in result analysis. Note that this composite approach still results in a single student model, by taking advantage of all the embeddings aforementioned.

\textbf{Extension to other networks}. We extend the aforementioned from the conventional TDNN-based model onto other x-vector variants, whose performances are reportedly better than their predecessor on established datasets. Three backbone architectures are included in this extension: Extended TDNN \cite{Snyder_etdnn_2019} (denoted as \emph{ETDNN}), ECAPA-TDNN \cite{ecapa} (\emph{ECAPA}), and Densely-connected TDNN \cite{dtdnn} (\emph{DTDNN}). All three models share a common structure featuring a final frame-level TDNN layer before the pooling layer and two fully-connected layers after. Consequently, we extract \emph{utterance} embeddings (from the first fully-connected layer) and \emph{wideBN} embeddings (from the final TDNN layer) from each of the three networks. We analyze the performance gap of the corresponding trained student networks. The training data for the teacher networks aligns with that used for the baseline TDNN x-vector. The dimension of the embeddings and the architecture of the corresponding student models remain consistent with earlier sections, and we utilize the pre-trained versions of the three aforementioned models.

All training steps including training of the teacher x-vector speaker encoder, have been carried out on a single NVIDIA TITAN V GPU. The evaluation of student models have been done without the reliance of GPU nodes.

\subsection{Performance Measures}
\label{ssec:evaluation}
We report ASV performance through \emph{equal error rate} (EER) and \emph{minimum detection cost function} (minDCF) with target speaker prior $p_{\mathrm{tar}} = 0.01$ and detection costs $C_\text{fa} = C_\text{miss} = 1.0$. We also show the \emph{detection error trade-off} (DET) curve for DNN-based systems for further analysis.

\begin{figure}[h]
    \centering
    \includegraphics[width=\linewidth]{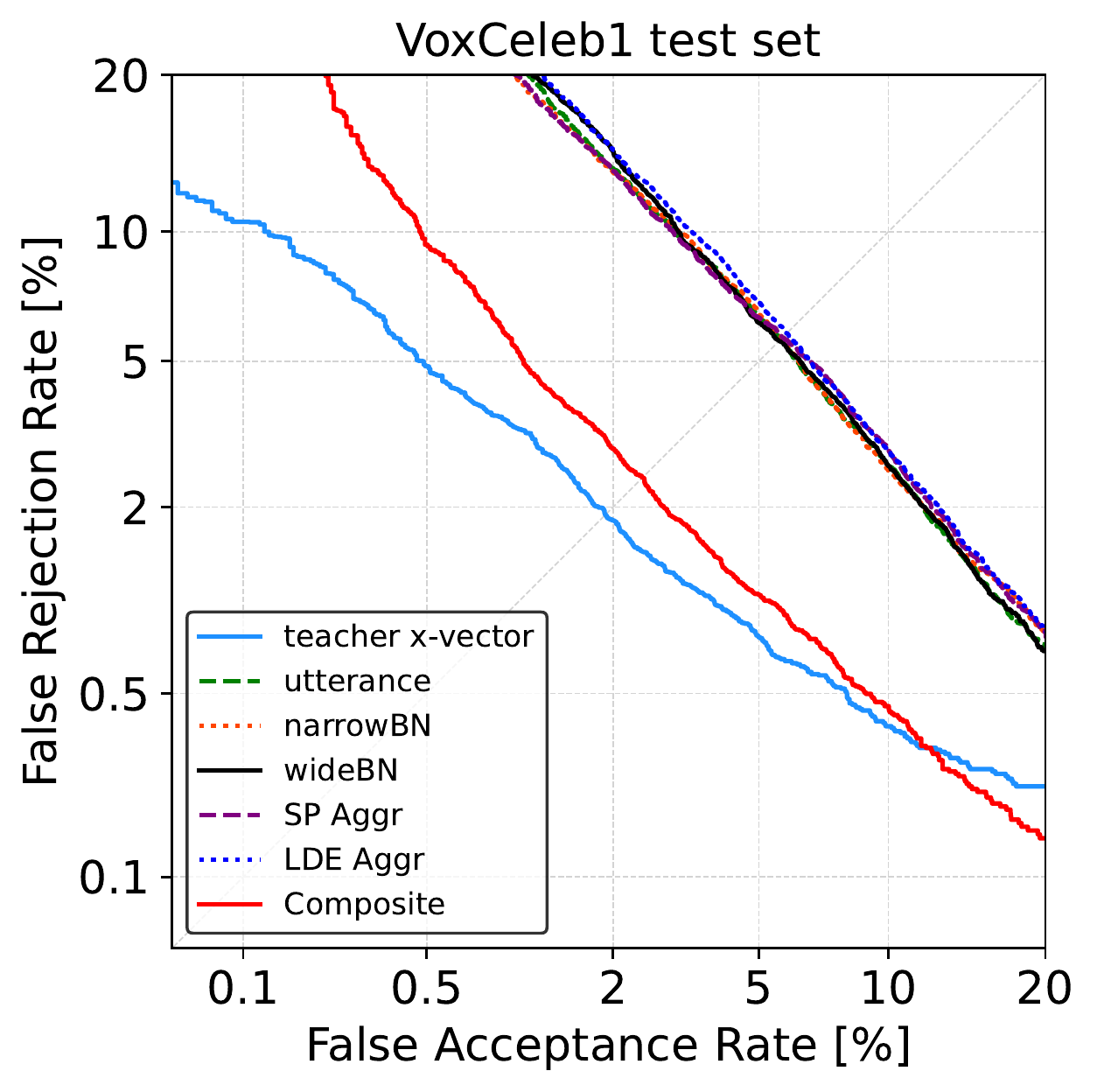}
    \caption{The DET profiles for the teacher x-vector and students, along with the composite system. The composite system here is without mean normalization. Best viewed in color.}
\label{fig:det_curve}
\end{figure}

\begin{figure}[ht]
    \centering
    \includegraphics[width=\linewidth]{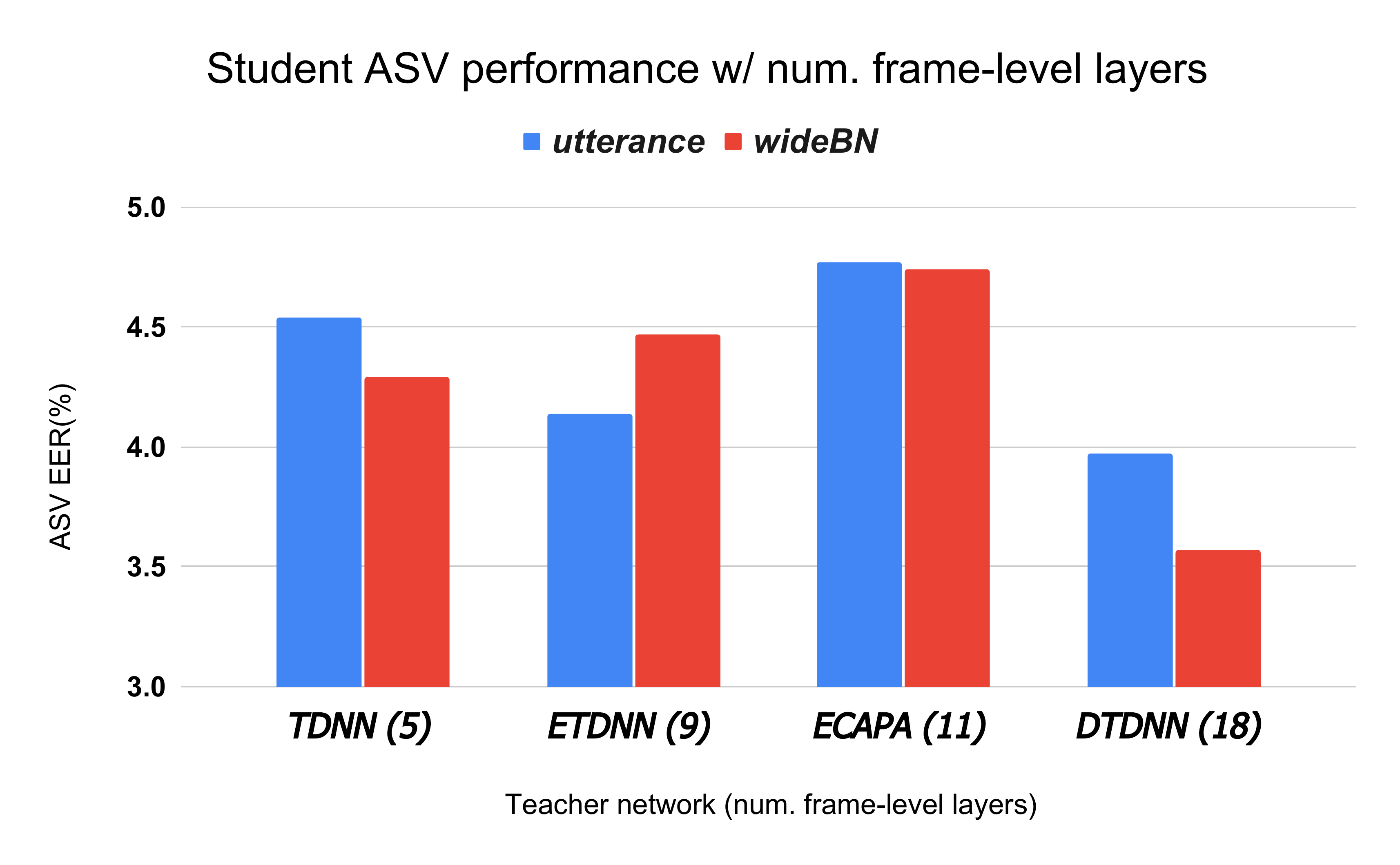}
    \caption{The relationship between the deepness of teacher models and student ASV performance. Best viewed in color.}
\label{fig:frame_asv_performance}
\end{figure}

\begin{table}[t] 
    \centering
    \small
    \caption{Results on VoxCeleb1 test set, including composite system results via concatenation of teacher embeddings. Number of parameters of the student model is indicated in the brackets. The baseline (teacher) x-vector has 5.90M parameters, and reached 1.88\%/0.152 on EER/minDCF.}
    \begin{tabular}{|c|c|c|c|}
    \hline
         Embedding & mean norm. & EER(\%) & minDCF \\ \hline
         \emph{utterance}~(0.54M) & --- & 4.54 & 0.463 \\ \hline
         \multirow{2}{7.25em}{\emph{narrowBN}~(0.54M)} & no & 4.29 & 0.464  \\\cline{2-4}
         & yes & 4.52 & 0.466 \\ \cline{1-4}
         \multirow{2}{6.75em}{\emph{wideBN}~(0.81M)} & no & 4.29 & 0.460 \\\cline{2-4}
         & yes & 4.50 & 0.469 \\ \cline{1-4}
         \emph{SP Aggr}~(0.68M) & --- & 4.17 & 0.407 \\ \hline
         \emph{LDE Aggr}~(0.54M) & --- & 5.86 & 0.534 \\ \hline \hline 
         \multirow{2}{7.75em}{Composite~(1.50M)} & no & \textbf{2.04} & \textbf{0.252}  \\\cline{2-4}
         & yes & 2.26 & 0.280  \\ \cline{1-4}
    \end{tabular}
    \label{tab:student_results}
\end{table}

\begin{table}[t]
    \centering
    \small
    \caption{Comparison with previous research on small footprint speaker verification. The cosine backend corresponds to the cosine scoring scheme \cite{cosine_similarity}. Their teacher ECAPA-TDNNs reached 0.99\%/0.113 \cite{lightweight1} and 1.14\%/0.075 \cite{lightweight2} on EER/minDCF, respectively.}
    \begin{tabular}{|c|c|c|c|c|}
    \hline
          & Large network & Backend & EER(\%) & minDCF \\ \hline
        \cite{lightweight1} & ECAPA-TDNN \cite{ecapa}& Cosine & 2.31 & \textbf{0.251} \\ \hline 
        \cite{lightweight2} & ECAPA-TDNN \cite{ecapa}& Cosine & 2.62 & 0.252 \\ \hline
        Ours & TDNN \cite{xvector2018} & PLDA & \textbf{2.04} & 0.252 \\ \hline
    \end{tabular}
    \label{tab:comparison_results}
\end{table}

\begin{table}[t]
    \centering
    \small
    \caption{Results on \emph{utterance} and \emph{wideBN} embeddings from different speaker encoders. Number of parameters of the teacher model is in the brackets. \emph{wideBN} Embeddings covered here do not apply mean normalization.}
    \begin{tabular}{|c|c|c|c|}
    \hline
        Teacher & embedding & EER(\%) & minDCF \\ \hline
        \multirow{2}{8.75em}{\emph{TDNN}~\cite{xvector2018}(5.90M) \\ {\footnotesize EER=1.88\%}} & \emph{utterance} & 4.54 & 0.463 \\ \cline{2-4}
        & \emph{wideBN} & 4.29 & 0.460  \\ \hline
        \multirow{2}{8.75em}{\emph{ETDNN}~\cite{Snyder_etdnn_2019}(13.2M) \\ {\footnotesize EER=1.67\%}} & \emph{utterance} & 4.14 & 0.462 \\ \cline{2-4}
        & \emph{wideBN} & 4.47 & 0.463  \\ \hline
        \multirow{2}{8.75em}{\emph{ECAPA}~\cite{ecapa}(6.35M) \\ {\footnotesize EER=1.66\%}} & \emph{utterance} & 4.77 & 0.484 \\ \cline{2-4}
        & \emph{wideBN} & 4.74 & 0.469  \\ \hline
        \multirow{2}{8.75em}{\emph{DTDNN}~\cite{dtdnn}(4.43M) \\ {\footnotesize EER=1.74\%}} & \emph{utterance} & 3.97 & 0.481 \\ \cline{2-4}
        & \emph{wideBN} & \textbf{3.57} & \textbf{0.455}  \\ \hline
    \end{tabular}
    \label{tab:network_extension}
\end{table}

\section{Results}
\label{sec:results}
Results for the baseline teacher and student systems along with their corresponding model sizes are presented in Table \ref{tab:student_results}.  sizes, are presented in Table \ref{tab:student_results}. The alternatives proposed in this work demonstrate comparable performance with utterance-level embeddings, with the exception of \emph{LDE Aggr}. Interestingly, the two frame-level embeddings outperform others when used without mean normalization. Notably, \emph{wideBN} without mean normalization outperforms \emph{utterance}, while \emph{narrowBN} without mean normalization yields equal EER performance with \emph{wideBN}, despite their different vector dimensionalities.
Among the single embeddings, \emph{SP Aggr} achieves the lowest EER and minDCF, while \emph{LDE Aggr} results in the highest values in both categories.

Results of the two composite systems are shown at the bottom of Table \ref{tab:student_results}. Both trained student networks outperform the single systems and significantly reduce the performance gap between student and teacher on both metrics, while still utilizing a relatively 75\% gap in terms of model size. This further capitalizes the usefulness of the enlisted embeddings extracted from the teacher, containing information that may have been missed by utterance-level embeddings. Meanwhile, deeper architectures with more resorting effort into engineering may result in better results, but that will lead to a larger size of the model. 

The comparison with previous works on small footprint models is outlined in Table~\ref{tab:comparison_results}. Although their final models are smaller, it is important to note that their teacher models are more robust, and their losses \cite{aam_softmax} are tailored for cosine similarity. Additionally, these approaches necessitates joint optimization of the teacher network, and the design of the student network can be intricate. In contrast, the proposed system, prioritizing simplicity, achieves competitive performance on both metrics. Meanwhile, it is worth noting that these performance figures may benefit from PLDA.

Fig.~\ref{fig:det_curve} presents the DET profile of the teacher x-vector and the described student models, including the two composite systems. While the single student system do not reach the teacher, the composite system demonstrates comparable performance.

Results of the experiments on the extension to other networks have been presented in Table~\ref{tab:network_extension}. 
The best single system performance is achieved by \emph{DTDNN} in terms of both metrics, utilizing its \emph{wideBN} embeddings. This is interesting as it has the fewest number of parameters among the models listed. The ``deepness" of the model, indicated by the number of frame-level layers, thus may positively influence the performance of the two types of embeddings. Fig.~\ref{fig:frame_asv_performance} depicts such relationships. 
While the performance relative to the number of frame-level layers is not monotonic, \emph{DTDNN}, with a notably deeper frame-level architecture than others, outperforms all other teachers. Future work may study on more effective ways to aggregate frame-level information from neural networks with deep architectures.

\section{Conclusion}
\label{sec:conclusion}
In this work, we have addressed the development of small footprint ASV systems through knowledge distillation. We have highlighted the potential of various types of embeddings extracted from the teacher x-vector network, encompassing both utterance-level and frame-level speaker information. \textbf{Our results on the standard VoxCeleb corpus confirm our hypothesis on the usefulness of multiple levels of speaker representation in small footprint ASV tasks.} 
The student systems consistently achieve an 85\%-91\% reduction in model size compared to the teacher model. Moreover, the composite student learner via concatenation of various types of speaker embeddings across multiple layers significantly outperforms individual models, reaching comparable performance with the teacher (2.04\% vs. 1.88\%), while still maintaining a substantial 75\% relative size reduction. We have extended our learning framework to incorporate several other teacher x-vector networks such as ECAPA-TDNN, revealing the effectiveness of deeper frame-level architectures and generalizability of our main findings. Future work may investigate on advanced aggregation methods and explore task-specific speaker embedding vectors, with the involvement of more advanced learning schemes, along with extension to stronger and more versatile teacher learning model.



 \bibliographystyle{elsarticle-num} 
 \bibliography{cas-refs}





\end{document}